\documentclass[]{aa}

\def\rmit#1{{\it #1}}              
\def\specchar#1{{\sc #1}}
\def\FeI{\mbox{Fe\,\specchar{i}}}

\def\CaIIH{\mbox{Ca\,\specchar{ii}\,\,H}}       

\def\ie{\rmit{i.e.}}
\def\eg{\rmit{e.g.}}

\def\pun{\stackrel{}{\mbox{.}}}
\def\farcs{$\stackrel{\prime\prime}{\pun}$}

\usepackage{booktabs}
\usepackage{graphicx}
\usepackage{multirow}
\usepackage{color}
\usepackage[flushleft]{threeparttable}

\usepackage{array}
\newcolumntype{?}{@{\vrule width 2pt}}

\usepackage{hyperref}
\hypersetup{
    final=true,
    pageanchor=true,
    colorlinks=true,
    breaklinks=true,
    linkcolor=blue,
    citecolor=blue,
    urlcolor=blue,
    pdfpagemode=UseNone,
    pdftitle={Signatures of the impact of flare ejected plasma on the photosphere of a sunspot light-bridge},
    pdfauthor={T. Felipe et al.},
    pdfsubject={Solar Physics},
    pdfkeywords={Methods: observational - methods: data analysis - Sun: photosphere - Sun: magnetic fields - sunspots - Sun: activity}}

\titlerunning{Signatures of flare ejected plasma on a light-bridge}   

\begin{document}


\title{Signatures of the impact of flare ejected plasma on the photosphere of a sunspot light-bridge}

\author{T. Felipe\inst{\ref{inst1},\ref{inst2}}
\and M. Collados\inst{\ref{inst1},\ref{inst2}}
\and E. Khomenko\inst{\ref{inst1},\ref{inst2}}
\and S. P. Rajaguru\inst{\ref{inst5}}
\and M. Franz\inst{\ref{inst4}}
\and C. Kuckein\inst{\ref{inst3}}
\and A. Asensio Ramos\inst{\ref{inst1},\ref{inst2}}
}


\institute{Instituto de Astrof\'{\i}sica de Canarias, 38205, C/ V\'{\i}a L{\'a}ctea, s/n, La Laguna, Tenerife, Spain\label{inst1}
\and 
Departamento de Astrof\'{\i}sica, Universidad de La Laguna, 38205, La Laguna, Tenerife, Spain\label{inst2} 
\and 
Indian Institute of Astrophysics, Bangalore-34, India \label{inst5} 
\and
Kiepenheuer-Institut f\"ur Sonnenphysik, Sch\"oneckstr. 6, 79104 Freiburg, Germany\label{inst4}
\and 
Leibniz-Institut f\"ur Astrophysik Potsdam (AIP), An der Sternwarte 16, 14482 Potsdam, Germany \label{inst3}
}

\abstract
{} 
{We investigate the properties of a sunspot light-bridge, focusing on the changes produced by the impact of a plasma blob ejected from a C-class flare.}
{We observed a sunspot in active region NOAA 12544 using spectropolarimetric raster maps of the four \FeI\ lines around 15655 \AA\ with the GREGOR Infrared Spectrograph (GRIS), narrow-band intensity images sampling the \FeI\ 6173 \AA\ line with the GREGOR Fabry-P\'erot Interferometer (GFPI), and intensity broad band images in G-band and \CaIIH\ band with the High-resolution Fast Imager (HiFI). All these instruments are located at the GREGOR telescope at the Observatorio del Teide, Tenerife, Spain. The data cover the time before, during, and after the flare event. The analysis is complemented with Atmospheric Imaging Assembly (AIA) and Helioseismic and Magnetic Imager (HMI) data from the Solar Dynamics Observatory (SDO). The physical parameters of the atmosphere at differents heights were inferred using spectral-line inversion techniques.}
{We identify photospheric and chromospheric brightenings, heating events, and changes in the Stokes profiles associated to the flare eruption and the subsequent arrival of the plasma blob to the light bridge, after traveling along an active region loop.}
{The measurements suggest that these phenomena are the result of reconnection events driven by the interaction of the plasma blob with the magnetic field topology of the light bridge.}


\maketitle


\section{Introduction}

The umbrae of sunspots usually exhibit bright elongated structures known as light bridges, especially during the fragmentation phase of a decaying sunspot or in the assembling process of magnetized regions that leads to the formation of a new spot. Light bridges have been classified according to their morphological properties \citep{Sobotka+etal1993, Sobotka+etal1994}, although all of them share common characteristics, such as their weaker and more horizontal magnetic field in comparison with the surrounding umbra \citep{Lites+etal1991, Leka1997}. The field lines around light bridges form a canopy structure \citep{Jurcak+etal2006, Lagg+etal2014} with reduced field strength in the inner part. This fact supports the scenario in which light bridges are created by the intrusion of field-free (or more specifically, weak-field) plasma into the strongly magnetized sunspot umbra \citep{Spruit+Scharmer2006, Schussler+Vogler2006} through vigorous convection \citep{Rimmele1997, RouppevanderVoort+etal2010,Lagg+etal2014}. 

Many dynamic phenomena have been reported in the chromosphere above light bridges \citep[\eg,][]{Roy1973, Asai+etal2001, Berger+Berdyugina2003, Shimizu+etal2009, Robustini+etal2016}. These events are driven by the interaction of the light-bridge magnetic topology with changes in the surrounding atmosphere produced by magnetoconvective motions. 
\citet{Toriumi+etal2015a} detected recurrent brightenings in AIA 1600 and 1700 \AA, Interface Region Imaging Spectrograph (IRIS) 1330 and 1400 \AA, and \CaIIH\ channels and dark surge ejections in AIA EUV data. The authors associated these brightenings to magnetic reconnection between the umbral vertical field and the emerged horizontal field to the surface from the convective upflow. Recently, \citet{Felipe+etal2016b} reported on the presence of magnetic field reversals in the photosphere of a light bridge. They appear due to the dragging of magnetic field lines by the convective motions in weak field regions, where the gas pressure is higher than the magnetic pressure. This magnetic field configuration can also trigger dynamic events at higher layers by means of reconnection of the reversed field lines.

Flares are one of the most energetic solar events. They are triggered via magnetic reconnection in the corona, and the subsequent downward energy transport can result in changes of the magnetic topology at the chromosphere and even at the photosphere \citep{Kosovichev+Zharkova1999, Sudol+Harvey2005, Petrie+Sudol2010, Fischer+etal2012}. An enhancement of the photospheric horizontal magnetic field has been detected at the flaring magnetic polarity inversion line after the eruption \citep{Wang+etal2012, Gomory+etal2017}, which confirms previous theoretical predictions \citep{Hudson+etal2008}. However, \citet{Kuckein+etal2015} found that during the peak of the flare the magnetic field (vertical and horizontal components) almost vanishes, and afterwards recovers its initial pre-flare configuration.  

The energy release in flares is sometimes accompanied by plasmoid ejections \citep[\eg,][]{Shibata+etal1995, Kim+etal2005}. These ejections have significant effect on the surrounding regions, such as originating extreme-ultraviolet wavefronts \citep{Kumar+Manoharan2013}. In case of low energy flares, plasmoids can be confined within active region loops. \citet{Zacharias+etal2011} investigated the ejection and travel of a trapped plasma blob using numerial simulations, while \citet{Yang+etal2016} analyzed the brightenings and height changes in a light wall (rooted in a light bridge) produced by the action of falling material ejected from a nearby flare. Recently, \citet{Nistico+etal2017} evaluated the impact of confined plasma blobs on magnetodydrodynamic waves studied in coronal seismology. 

As a result of a lucky coincidence, we were able to acquire spectropolarimetric data of a sunspot light-bridge prior to the occurrence of a C-class flare in the same active region and after the arrival of a plasma blob ejected by that flare. We aim to evaluate the effects of this event on the dynamics and magnetic topology of the light-bridge atmosphere. The organization of the paper is as follows: Sect. \ref{sect:observations} presents the data and the analysis methods, Sect. \ref{sect:global} shows the general characteristics of the event under study, Sect. \ref{sect:configuration} describes the properties and evolution of the light-bridge structure, and Sect. \ref{sect:stokes} analyze the changes in the Stokes profiles during the observations. Finally, discussion and conclusions are given in Sects. \ref{sect:discussion} and \ref{sect:conclusions}.

\section{Observations}
\label{sect:observations}

\subsection{Data set}
\label{sect:data}

We focus on the analysis of the sunspot in active region NOAA 12544. The data were obtained on 2016 May 15. The active region emerged at the north hemisphere of the Sun a few days prior to that date. In this work we combine data from multiple instruments located at ground-based and space-based observatories.

\begin{figure}[!ht] 
 \centering
 \includegraphics[width=9cm]{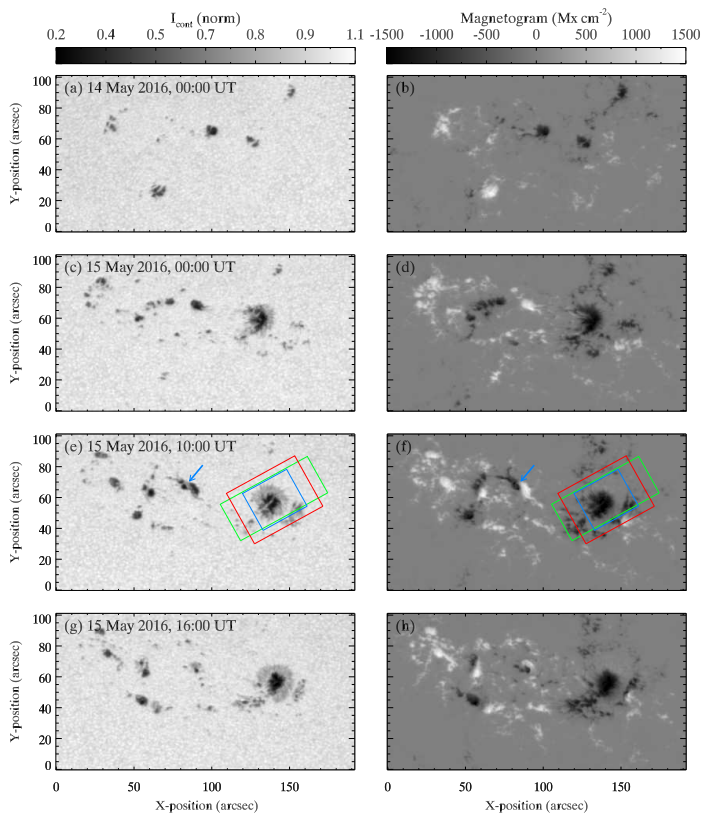}
  \caption{HMI continuum images (left column) and magnetograms (right column) of the evolution of active region NOAA 12544. Date and time of the images are indicated at the top of the continuum images. The colored rectangles in panels (e) and (f) indicate the FOV covered by GRIS (green), GFPI (red), and HiFI (blue). The blue arrow indicates the location of a C1.3-class flare during the observations. See movie in the online material.}
  \label{fig:evolution_HMI}
\end{figure}

Spectropolarimetric raster maps of the sunspot were scanned using the GREGOR Infrared Spectrograph \citep[GRIS,][]{Collados+etal2012}, installed at the 1.5-meter GREGOR solar telescope \citep{Schmidt+etal2012} at the Observatorio del Teide, Tenerife, Spain. Three maps with the four Stokes parameters were acquired between 09:41 UT and 10:10 UT on 2016 May 15. The slit covers $63.''5$ with a pixel size of $0.135''$. For the first two maps 200 slit positions were acquired in the raster scanning with a step size of $0.135''$. The last map was larger, spanning 300 slit positions with the same step size. The dimensions of the common field-of-view (FOV) were $63.''5 \times 27''$. The FOV is delimited by a green line in panels (e) and (f) from Fig. \ref{fig:evolution_HMI}. Three accumulations with an integration time of 100 ms were taken at each scan. The whole FOV was scanned in eight minutes. The spectral region covers a wavelength range of \hbox{40 \AA} around \hbox{15655 \AA} with a wavelength sampling of \hbox{40 m\AA\ pixel$^{-1}$}. This region includes several photospheric \FeI\ spectral lines with high magnetic sensitivity (see Sect. \ref{sect:GRIS}). The GREGOR adaptive optics system \citep[GAOS,][]{Berkefeld+etal2012} was performing well during the acquisition of the data. Polarimetric calibration as well as dark and flat-field corrections were applied following the standard procedures described in \citet{Collados1999, Collados2003}. See \citet{Franz+etal2016} for a detailed study on the spectral quality of the GRIS data.

Two-dimensional imaging spectroscopy sampling the photospheric \hbox{\FeI\ 6173 \AA} line was obtained with the GREGOR Fabry-P\'erot Interferometer \citep[GFPI,][]{Puschmann+etal2012}. The series started at 08:58 UT and finished at 10:02 UT, May 15. The image scale in the $2\times2$-pixel binning mode was 0\farcs081\,pixel$^{-1}$. The images had a size of $688 \times 512$ pixel which translated into a FOV of $55.''7\times 41.''5$ (see red rectangle in panels (e) and (f) from Fig. \ref{fig:evolution_HMI}). The intensity profile of the \hbox{\FeI\ 6173 \AA} line was sampled with 32 wavelength points, providing a temporal cadence of 27.7 s. The data were processed using the data reduction pipeline ``sTools'' \citep{Kuckein+etal2017}, and images were restored with Multi-Object Multi-Frame Blind Deconvolution \citep{Lofdahl2002, vanNoort+etal2005}. The same time range was observed with the High-resolution Fast Imager (HiFI, C. Denker, private communication). The instrument consists of two synchronized sCMOS cameras which are located in the blue imaging channel of the GFPI. Data were simultaneously taken using a G-band (4307 \AA) and a \CaIIH\ band (3968 \AA) filter. The exposure time was 2 ms. HiFI acquires data at a very high frame rate, and only the best 100 images out of 500 are used for image restoration using KISIP \citep{Woger+vonderLuhe2008}. The spatial sampling is $0.''0253$ pixel$^{-1}$ and the FOV is $48.''6\times 27.''3$ (see blue rectangle in panels (e) and (f) from Fig. \ref{fig:evolution_HMI}).

The ground-based observations obtained from the GREGOR telescope were complemented with data from the Solar Dynamics Observatory \citep[SDO,][]{Pesnell+etal2012}. The long-term photospheric evolution was examined using 12-minute intensity images and magnetograms from the Helioseismic and Magnetic Imager \citep[HMI,][]{Schou+etal2012}. The pixel size was $0.''50$. We have selected a FOV of $190''\times100''$, which includes not only the sunspot observed with GREGOR, but also the whole active region. 
The extreme ultraviolet bands (EUV, 94, 131, 171, 193, 211, 304, and 335 \AA) from the Atmospheric Imaging Assembly \citep[AIA,][]{Lemen+etal2012} were analyzed during the ground-based observing time (09:41 -- 10:10 UT). The temporal cadence of the EUV channels was 12 s whereas the ultraviolet bands (1600 and 1700 A) had a longer cadence of 24 s. HMI and AIA data were resampled to a common image scale of $0.''6$ pixel$^{-1}$ using the routine $aia\_prep$ from the $SolarSoftWare$ package \citep{Freeland+Handy1998}. 

The coalignment between different instruments was obtained by looking for the spatial shift and rotation which yield the best correlation between cotemporal photospheric images from each data set. For the SDO data the HMI continuum intensity was used as a reference. For GRIS and GFPI data we also selected continuum images, while in the case of HiFI the G-band data was selected as a reference.

\subsection{GRIS data and spectral line inversions}
\label{sect:GRIS}

The Stokes parameters observed from ground-based telescopes at each spatial position differ from those emitted on the solar surface due to the turbulence in the Earth atmosphere and the optical properties of the instrument. These effects can be quantified through the point spread function (PSF) and the spatial resolution of the data can be improved by performing the spatial deconvolution of the observed data. For the analysis of solar images acquired as two-dimensional data, a dynamic PSF can be determined empirically using reconstruction techniques \citep[\eg,][]{vonderLuhe1993, Lofdahl2002, vanNoort+etal2005}, as mentioned above for the GFPI observations. These methods account for the temporal variation of the seeing. In the case of long-slit spectrograph, such as GRIS, these techniques are not applicable and obtaining a PSF that characterizes the optical system, including the seeing, remains a challenge \citep{Beck+etal2011}.  

We have deconvolved the GRIS maps using principal component analysis regularization \citep{RuizCobo+AsensioRamos2013, QuinteroNoda+etal2015}. We use the empirically determined PSF of GRIS from the observations of the Mercury transit on 2016 May 9. It comprises two different contributions, one of them accounting for the instrumental stray light and the other for the resolution of the telescope. The latter includes the seeing, which is estimated from the average value of the Fried parameter $r_0$ during the scanning of each map. In this data set, the average $r_0$ at 15650 \AA\ was $32$ cm. Details of the empirical PSF can be found in \citet{Felipe+etal2016b}.

We have analyzed the photospheric \FeI\ spectral lines at 15648.5, 15652.9, 15662.0 and 15665.2 \AA. The atomic parameters are summarized in Table \ref{table:atomic_data}. The central laboratory wavelengths, excitation potentials, and electronic configurations were obtained from \citet{Nave+etal1994}. The parameters $\sigma$ and $\alpha$, which describe the line broadening due to collisions with neutral atoms, were estimated under the ABO theory \citep{Anstee+O'Mara1995} and were taken from \citet{Borrero+etal2003} for the first two lines and from \citet{Bloomfield+etal2007a} for the other two lines. The oscillator strengths correspond to empirical calculations and were taken from \citet{Borrero+etal2003} and \citet{Bloomfield+etal2007a}. The \FeI\ 15648.5 \AA\ line has a large Land\'e factor and has high sensitivity to the magnetic field, while the \FeI\ 15652.9 and \hbox{15662.0 \AA} lines are moderately magnetically sensitive. We use the solar abundances from \citet{Asplund+etal2009}.

\begin{table*}
\begin{center}
\caption[]{\label{table:atomic_data}
          {Atomic parameters of the observed lines.}}
\begin{tabular*}{12.5cm}{ cccccccc}
\hline\noalign{\smallskip}
Ion & 	$\lambda_{\rm 0}$ (\AA)	& $\chi_{\rm l}$	& $\log(gf)$	& Elec. configurations	& $\sigma$	& $\alpha$ & $\bar{g}$	\\
\hline\noalign{\smallskip}
\FeI\ & 15648.515		& 5.426			& -0.739		& $^{7}D_1- ^{7}D_1$	& 975		& 0.229 	  & 3.00	\\
\FeI\ & 15652.874 		& 6.246			& -0.165		& $^{7}D_5- ^{6}D_{4.5}4f[3.5]^0$	& 1427		& 0.330 	  & 1.45	\\
\FeI\ & 15662.018		& 5.828			& 0.120			& $^{5}F_5- ^{5}F_4$	& 1200		& 0.24 	  & 1.50	\\
\FeI\ & 15665.245		& 5.979			& -0.490		& $^{5}F_1- ^{5}D_1$	& 1283  	& 0.23	  & 0.75	\\ 	
\FeI\ & 6173.336		& 2.223			& -2.919		& $^{5}P_1- ^{5}D_0$	& 281  	& 0.27	  & 2.50	\\

\hline

\end{tabular*}

\begin{tablenotes}
\small
 \item The first column indicates the ion, $\lambda_{\rm 0}$ is the laboratory wavelength,  $\chi_{\rm l}$ is the excitation potential of the lower energy level in eV, $\log(gf)$ is the logarithm of the oscillator strength times the multiplicity of the level, $\sigma$ (in units of the Bohr radius squared, $a_0^2$) and $\alpha$ are the collisional broadening parameters, and $\bar{g}$ is the effective Land\'e factor. The tabulated values of $\log(gf)$ correspond to a solar iron abundance of 7.50 dex \citep{Asplund+etal2009}.

\end{tablenotes}
 
\end{center}
\end{table*}

The four spectral lines were inverted simultaneously using the SIR (Stokes Inversion based on Response functions) code \citep{RuizCobo+delToroIniesta1992}, which provides the stratification with the continuum optical depth of the atmospheric parameters that best reproduce the observed Stokes profiles. We have used the cool umbral model from \citet{Collados+etal1994} as initial guess model. The temperature stratification was iterativelly modified with five nodes, while two nodes were used for the rest of the physical parameters (line-of-sight (LOS) velocity, magnetic field strength, magnetic field inclination and azimuth, and microturbulence). The magnetic field has been converted to the local solar reference frame, where the vertical magnetic field is radially oriented.

The 180$^{\circ}$ ambiguity in the azimuth retrieved from the inversions has been resolved using the method from \citet{Leka+etal2009proc}\footnote{publicly available at http://www.cora.nwra.com/AMBIG/}. In this code the ambiguity is first resolved based on the minimum energy method \citep{Metcalf1994}. Since this method can fail in the presence of noise \citep{Leka+etal2009}, the pixels with a transverse field strength below a selected threshold are then recalculated using an iterative acute-angle criteria to surrounding positions following \citet{Canfield+etal1993}.

\section{Global properties}
\label{sect:global}

\subsection{Photospheric evolution of the active region}
\label{sect:evolution}

Figure \ref{fig:evolution_HMI} illustrates the temporal evolution of active region NOAA 12544 as seen in HMI continuum intensity and magnetograms over two days around the date of the GREGOR observations. A movie showing the long-term evolution over seven days can be found in the online material. 

The active region rotated into the visible solar hemisphere as a few pores. Starting at 20:00 UT 2016 May 12, new flux emergence appeared at the east of the former pores. The magnetic flux with negative polarity was displaced towards the west, while the positive polarity moved towards the east. One of these recently emerged pores is located at the middle ($x\sim90''$, $y\sim 65''$) of panels (a) and (b) in Fig. \ref{fig:evolution_HMI}. It continues moving towards the west, and on May 14 at 15:00 UT it starts to develop a penumbra (see panel (c) from Fig. \ref{fig:evolution_HMI} at $x\sim 130''$, $y\sim 60''$). The umbral part changes its shape and surrounds an elongated region of penumbral-like atmosphere, forming the light bridge under study in this work. Panels (e) and (f) from Fig. \ref{fig:evolution_HMI} show the HMI data at the time the GREGOR observations were obtained, with the FOV of the used instruments indicated by colored rectangles. The light bridge completely divides the umbra and it has a length of $13.''5$ and a variable thickness that changes from approximately $2''$ at the east side to less than $1''$ at the west side. It also shows a smaller and thinner branch close to the eastern umbral boundary. A few hours later, the umbra at both sides of the light bridge merges (panels (g) and (h) from Fig. \ref{fig:evolution_HMI}). The life time of the light bridge is around seven hours. The recently formed sunspot continues its evolution by merging with some of the adjacent pores until it disappears across the western limb.

\subsection{C-class flare}
\label{sect:flare}

\begin{figure}[!ht] 
 \centering
 \includegraphics[width=9cm]{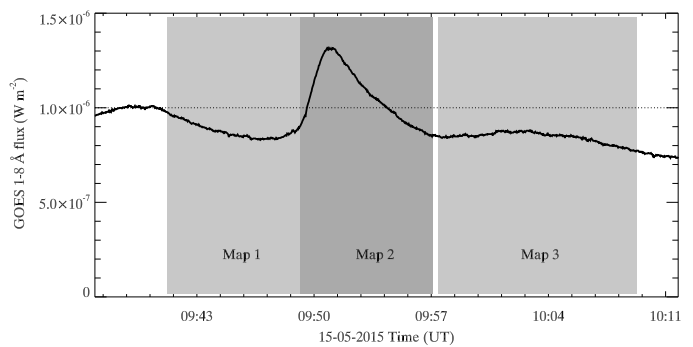}
  \caption{Temporal variation of GOES X-ray flux. The dotted horizontal line indicates the threshold for C-class flares. Grey shaded areas mark the time range of the three raster spectropolarimetric maps measured with GREGOR/GRIS.}
  \label{fig:GOES}
\end{figure}

A C1.3 flare took place at the same active region during the GREGOR observations. The Geostationary Operational Enviromental Satellite \hbox{(GOES) 1-8 \AA\ } X-ray flux (Fig. \ref{fig:GOES}) began to rise at 09:47 UT, very close to the start of the scanning of the second GRIS map. The peak was reached at 09:51 UT, and then it smoothly decreases until returning to pre-flare values at 09:57 UT. Maps 1, 2, and 3 scanned with GRIS captured the active region before the flare, during the flare, and just after the flare, respectively.

\begin{figure}[!ht] 
 \centering
 \includegraphics[width=9cm]{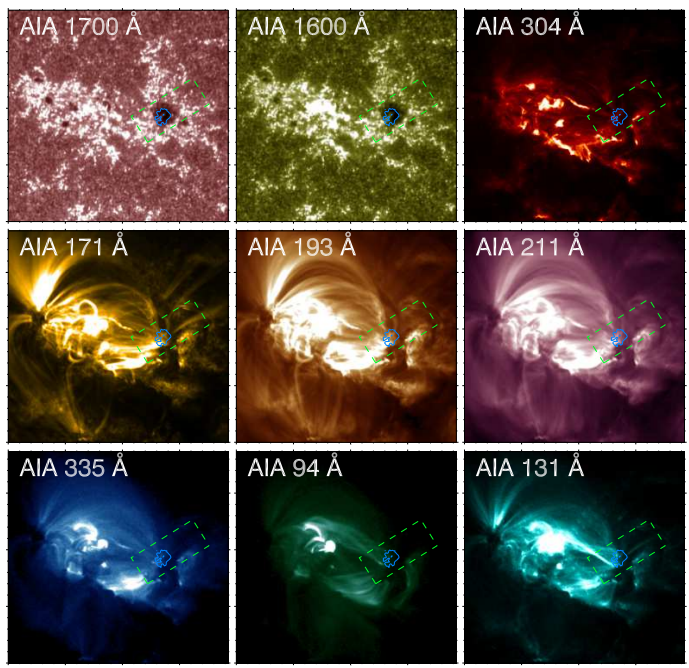}
  \caption{SDO/AIA filtergrams of active region NOAA 12544 at 09:51:42 UT. The band is indicated at the top-left corner or each panel. The dashed green line is the FOV of GREGOR/GRIS data. The solid blue line represents the contour of the umbra as seen in HMI continuum images. See movie in the online material.}
  \label{fig:AIA_movie}
\end{figure}

\begin{figure}[!ht] 
 \centering
 \includegraphics[width=9cm]{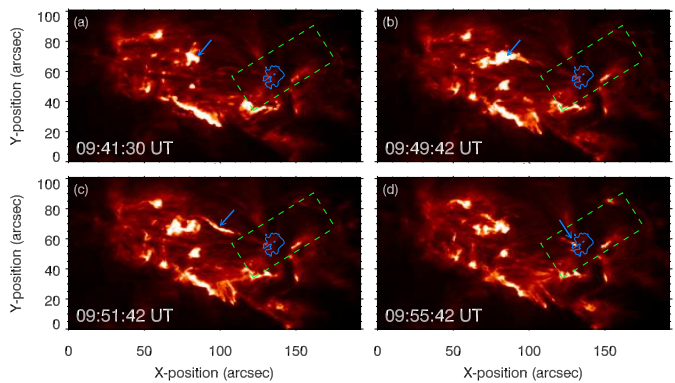}
  \caption{SDO/AIA 304 \AA\ filtergrams of active region NOAA 12544 at four time steps around the occurrence of a C1.3 flare. The dashed green line is the FOV of GREGOR/GRIS data. The solid blue line represents the contour of the umbra as seen in HMI continuum images. The blue arrow points to the origin of the flare in the top panels and to the position of the ejected plasma blob in the bottom panels.}
  \label{fig:AIA304A}
\end{figure}

Figure \ref{fig:AIA_movie} illustrates the active region recorded in all the SDO/AIA bands at 09:51 UT. The FOV from GRIS is outlined by the dashed green line, whereas the blue solid line marks the border of the umbra of the sunspot as seen in the HMI continuum images. The associated movie shows the temporal evolution of the flare event. Figure \ref{fig:AIA304A} shows several moments of this event in AIA 304 \AA. Panel (a) illustrates the image at the time the first GRIS scan started. A brightening is observed at the place marked by the blue arrow. The strongest emission at that location happened at 09:51 UT, which coincides with the peak of the C1.3 flare. The position coincides with the polarity inversion line which separates two small areas of opposite polarities. 
 (panel (f) from Fig. \ref{fig:evolution_HMI}). A bright blob of plasma departs from the flare location (panel (b) from Fig. \ref{fig:AIA304A}) and moves along a loop towards the sunspot situated at the west (blue arrow in panel c). The travel of the blob is visible in all EUV bands from AIA except for the two bands sensitive to the highest temperatures (94 and 131 \AA, Fig. \ref{fig:AIA_movie}). Only an increase in the emission of the loop is noticeable in these latter two filters. While the blob's trajectory is also apparent in AIA 1600 \AA\ filtergrams, which is sensitive to upper photospheric and transition region heights, the AIA 1700 \AA\ images showed no trace of it. No measurable time lag above the 12 s cadence of the AIA EUV data is found between the arrival of the brightening to a fix spatial position for any pair of EUV bands. At around 09:56 UT, the brightening reaches the umbra of the sunspot covered by our GREGOR observations at a location close to the light bridge.

This active region showed some significant activity around the time of the GREGOR observations, with the occurrence of 10 C-class flares in a time period of about 34 hours.

\section{Detailed configuration and evolution}
\label{sect:configuration}

\begin{figure}[!ht] 
 \centering
 \includegraphics[width=9cm]{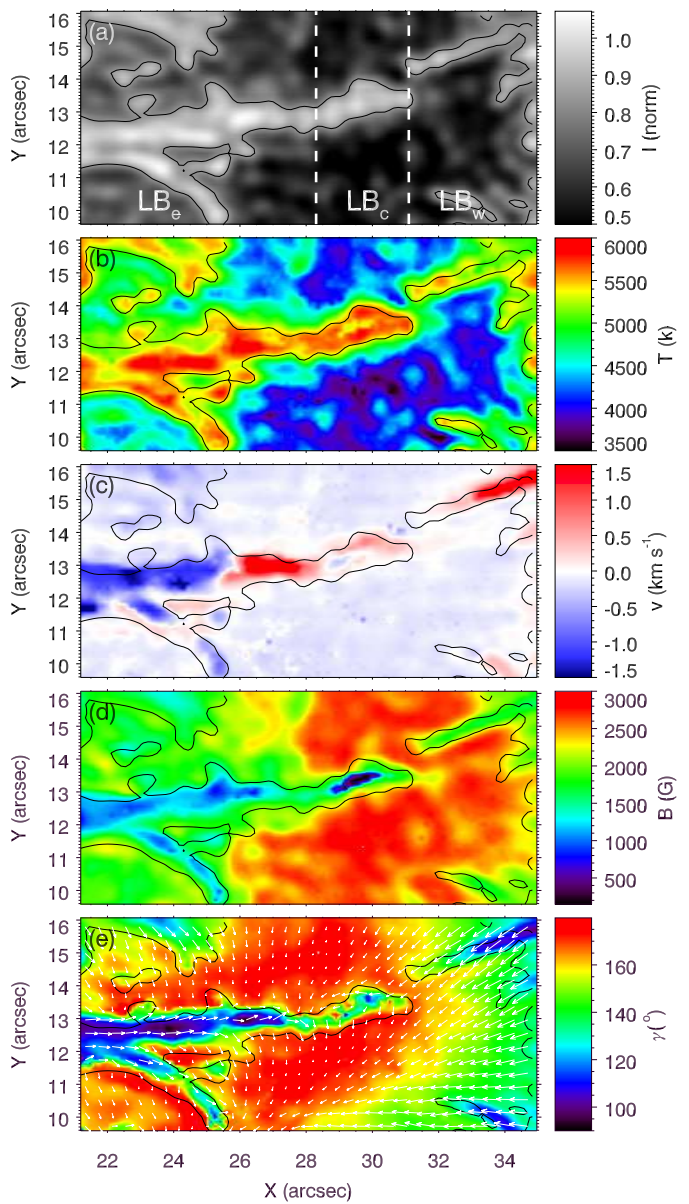}
  \caption{Maps of deconvolved continuum intensity (a), temperature (b), LOS velocity (c), magnetic field strength (d), and magnetic field inclination measured from the local radial direction (e) of the light bridge measured in the first scan (from 09:41 to 09:49 UT) of GREGOR/GRIS data. Panels b-e illustrate the results of simultaneous inversions of four infrared \FeI\ lines at $\log\tau=-0.5$ with SIR. White arrows in panel (e) indicate the orientation of the magnetic field in the horizontal plane, with their length proportional to the horizontal field strength. Black lines show contours of constant continuum intensity.}
  \label{fig:GRIS_inv}
\end{figure}

\begin{figure*}[!ht] 
 \centering
 \includegraphics[width=18cm]{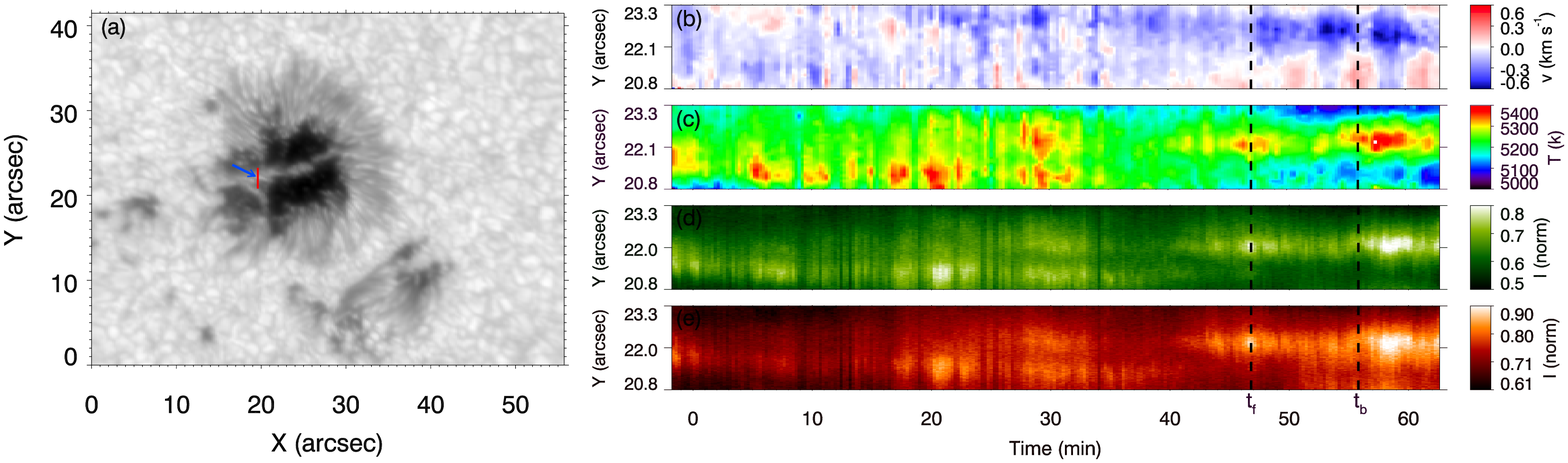}
  \caption{Left panel: Broad-band MOMFBD reconstructed image from GFPI at 09:58 UT. Right panels: Temporal evolution of the Doppler velocity (b) and temperature (c) at $\log\tau=-0.6$ inferred from the inversion of the \FeI\ 6173 \AA\ line (GFPI data), and the intensity evolution in G-band (d) and \CaIIH\ (e) from HiFI at the position indicated by a red line in panel (a). Time is indicated in minutes after 09:00 UT. The vertical dashed line denoted as $t_{\rm f}$ marks the time of the beginning of the flare and the dashed line $t_{\rm b}$ indicates the arrival of the plasma blob to the light bridge. The blue arrow in panel (a) points to the location of the spectra illustrated in Figs. \ref{fig:stokes} and \ref{fig:inversion}.}
  \label{fig:time_GFPI}
\end{figure*}

\subsection{Photospheric structure}

Figure \ref{fig:GRIS_inv} illustrates a detailed view of the continuum intensity in the region surrounding the light-bridge (panel a) and the magnetic and thermodynamic parameters retrieved from the SIR inversion (panels b-e). Inversion results are shown at an optical depth at 5000 \AA\ of $\log\tau=-0.5$. At this optical depth, the infrared \FeI\ lines used for the inversion have the peak in their sensitivity to magnetic field and velocity, which is deeper than that for other commonly used photospheric lines such as the \FeI\ pair at 6301 \AA\ \citep{Borrero+etal2016}. The figure shows the inversion obtained for the first GRIS map, but the results are qualitatively similar for the other two maps. 

The light bridge crosses all the umbra, except for a section at $X=31''$, $Y=14''$ where a small umbral atmosphere divides it. At the eastern part, a thin branch departs from the main light bridge. This part is surrounded by a faint umbra. Around the rest of the light bridge the umbra is darker in the continuum-intensity image. The light-bridge areas show some similar properties: (1) the continuum intensity is above approximately 85\% of the quiet-Sun intensity, (2) the temperature is higher than in the surrounding umbra, and (3) the magnetic field strength is reduced and its orientation becomes more horizontal in comparison with the umbral field.    

However, some of the properties show dependence with their position along the light bridge. We define three different parts of the light bridge as marked with vertical dashed lines in the top panel of Fig. \ref{fig:GRIS_inv}. The eastern part (LB$_{\rm e}$) includes the region with $X\lesssim 28''$ where the second branch is present. The central part (LB$_{\rm c}$) covers $28''\lesssim X \lesssim 31''$ and is separated from the western part (LB$_{\rm w}$) by the aforementioned light-bridge gap.

The umbra around the LB$_{\rm e}$ region shows a weaker field strength (between 1700 and 2300 G) and higher temperature (between 4500 and 5100 K) than the central part of the umbra. In the LB$_{\rm e}$ region the field strength decreases down to $\sim$1000 G and the temperature increases up to \hbox{6000 K}. The flow pattern in LB$_{\rm e}$ shows a large region with upflows (negative LOS velocities, blue color in panel c) and a downflow (positive LOS velocities, red color) closer to the center. The upflow is present in the main part of the light-bridge and also in the thin branch. Its maximum magnitude is \hbox{$-1.8$ km s$^{-1}$}. The maximum value of the downflow is \hbox{2.8 km s$^{-1}$}. The second and third GRIS maps show a similar flow pattern, but with clear differences in the amplitude of the flows. See Sect. \ref{sect:GFPI} for an examination of the changes in the flow velocities. As previously indicated, the magnetic field inclination in the light bridge is more horizontal than in the umbra. In the LB$_{\rm e}$ region the field is almost completely horizontal, with inclinations between 90$^{\circ}$ and 115$^{\circ}$. Interestingly, the most horizontal parts ($\gamma\sim 90^{\circ}$) coincide with the location of the stronger upflows and downflows. At the spatial position where the direction of the flow changes, the inclination is less horizontal ($\gamma\sim 115^{\circ}$). These flows are not oriented along the magnetic field files. Instead, they are embedded in an atmosphere dominated by the gas pressure. In the LB$_{\rm e}$ regions with absolute Doppler velocity higher than 0.5 km s$^{-1}$, the average plasma-$\beta$ (defined as the ratio between gas pressure and magnetic pressure) is 2.2. In this high-$\beta$ regime, plasma motions drag the field lines. In the main part of the LB$_{\rm e}$ region, the magnetic field (mostly horizontal) is directed along the axis of the light bridge, in the direction of increasing $X$. It is surrounded by vertical field, whose weak horizontal component is perpendicular to the one of the LB. Therefore, at the boundaries of the light bridge we find a $90^{\circ}$ change in the orientation of the horizontal field (at a constant optical depth) and an enhancement in the magnitude of the vertical component of the electric current density ($|j_z|\sim 0.5$ A m$^{-2}$). Note that at the thin branch the magnetic field is also oriented along the light bridge.

The central part of the light bridge (LB$_{\rm c}$) shows significantly lower values of the field strength, as low as $B\sim130$ G, with an inclination around $140^{\circ}$. The LOS velocity and the temperature at LB$_{\rm c}$ show evidence of convection. A central upflow with a velocity of $-0.3$ km s$^{-1}$ is surrounded by downflows ($\sim$0.7 km s$^{-1}$). The location of the upflow corresponds to a region of hot plasma at deeper layers around $\log\tau=0.4$ (not shown in the figure), where the response function of the infrared \FeI\ lines to the temperature is higher according to \citet{Borrero+etal2016}, while the downflows carry cooler material. The temperature at higher layers ($\log\tau=-0.5$, panel (a) from Fig. \ref{fig:GRIS_inv}) shows a central area with lower temperatures surrounded by higher temperatures. This pattern is similar to the reversed granulation  detected in photospheric spectral lines \citep[\eg,][]{Balthasar+etal1990, Janssen+Cauzzi2006}. It is due to the imbalance between the cooling produced by the adiabatic expansion of the fluid elements and the radiative heating, in combination with the higher vertical speed of the downflows in comparison with the central upflows \citep{Cheung+etal2007a}. These results confirm the vigorous convective nature of the LB$_{\rm c}$ region. 

The LB$_{\rm w}$ region is detached from the rest of the light bridge. This part is a penumbral filament that penetrates into the umbra. Typical penumbral values of the magnetic field strength ($\sim$1700 G) and field inclination ($100^{\circ}$) are present. Furthermore, as a result of the Evershed flow, we see downflows across the LB. We observe a gradient in the inferred redshifts starting with low values (around \hbox{2 km s$^{-1}$)} close to LB$_c$ and increasing towards the outer parts of the sunspot (see Fig. \ref{fig:GRIS_inv}c). This velocity is consistent with the Evershed flow in other penumbral filaments which do not penetrate into the umbra. The orientation of the horizontal magnetic field around LB$_{\rm w}$ is not modified due to the presence of the penumbral-filament-like light-bridge.

\subsection{Temporal evolution at the light bridge}
\label{sect:GFPI}

The intensity profiles of the \FeI\ 6173 \AA\ line obtained with the GFPI around the light bridge were also inverted using SIR. The atomic parameters of the line are included in Table \ref{table:atomic_data}. The inversions were performed with the same number of nodes as used for the inversion of the GRIS data, but an average light-bridge atmosphere (retrieved from the GRIS inversion) was employed as initial guess atmosphere. Although the magnetic field is also inferred from the inversion, since no Stokes $Q$, $U$, and $V$ parameters are present, we only focus on the LOS velocity and temperature from the inversions.

Figure \ref{fig:time_GFPI} illustrates the temporal evolution of several variables obtained from GFPI and HiFI at the red slit position shown in panel (a). Two vertical lines in panels (b)--(c) of Fig. 5 mark the beginning of the flare ($t_{\rm f}$=09:47 UT) and the collision of the plasma blob with the light bridge ($t_{\rm b}$=09:56 UT). The blue arrow in Fig. \ref{fig:time_GFPI}a indicates where the plasma blob collided with the light bridge. The light bridge is located at one of the footpoints of an EUV loop (see online movie). At the other side of the loop (outside the FOV of GFPI and GRIS) the flare starts at $t_{\rm f}$. This time coincided with the appearance of a brightening in G-band (panel d) and \CaIIH\ (panel e), and also with an increase of the photospheric temperature inferred from the \FeI\ 6173 \AA\ line (panel c). Later, a strong increase of the intensity is measured at both HiFI filters just after the plasma blob reaches the light bridge ($t_{\rm b}$). At the same location and time, the photospheric temperature at $\log\tau=-0.5$ increases by about 200 K. 

As described in Sect. \ref{sect:evolution}, we ascribe the formation of the light bridge to the approximation of the umbra at both sides of a penumbral-like region. The lifetime of the light bridge is only $\sim$7 hours, and during this time the light bridge is continously evolving. The GFPI broad-band images show that both sides of the umbra were getting closer, reducing the width of the light bridge. The Doppler velocity at the slit location illustrated in \hbox{Fig. \ref{fig:time_GFPI}} shows mainly blueshifts. The upflow increases towards the end of the time series. As the strongly magnetized umbral atmospheres approach each other, the plasma of the light bridge is confined to a narrower region and the upflows need to increase owing to mass conservation. After 09:40 UT the blueshifts show fluctuations with a period of about 5 min, which is consistent with waves excited by the global $p-$modes.

\section{Changes in Stokes profiles}
\label{sect:stokes}

\subsection{Description}

\begin{figure}[!ht] 
 \centering
 \includegraphics[width=9cm]{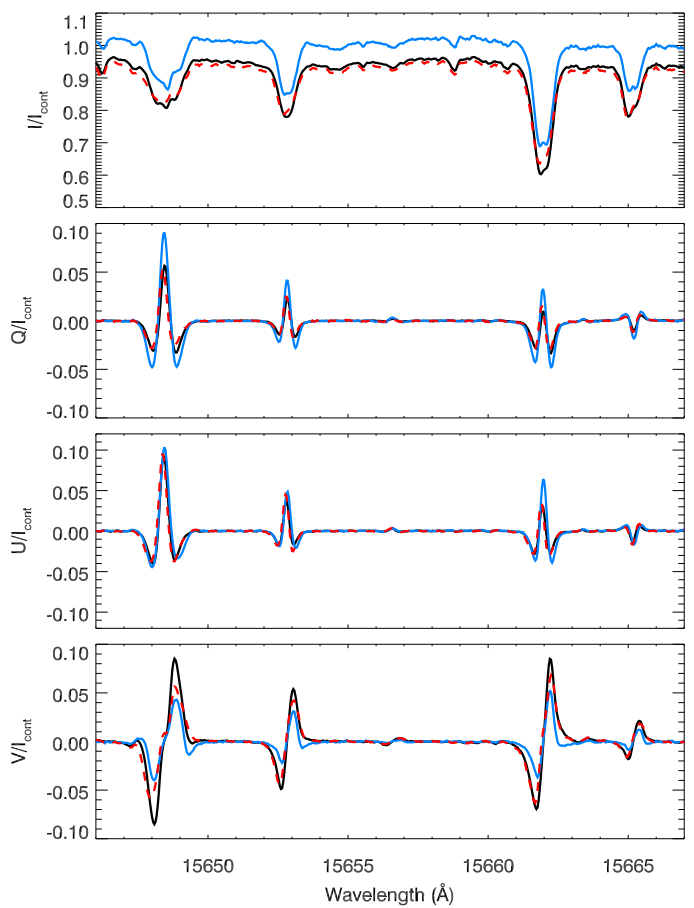}
  \caption{Stokes profiles of the four analyzed \FeI\ infrared lines at the position indicated by a blue arrow in panel (a) of Fig. \ref{fig:time_GFPI} for the first (black), second (dashed red), and third (blue) GRIS maps.}
  \label{fig:stokes}
\end{figure}

Figure \ref{fig:stokes} shows the four Stokes profiles measured with GRIS in the spectral region between 15646 and 15667 \AA. This spectral window includes the four infrared \FeI\ lines used for the inversion presented in Sect. \ref{sect:GRIS}. The figure illustrates three spectra corresponding to the three GRIS maps at the spatial position of the light bridge where a temperature increase was detected in the GFPI data (blue arrow in panel (a) from Fig. \ref{fig:time_GFPI}).

The intensity profile of the \FeI\ 15648.5 \AA\ line shows an irregular shape as a result of the partial separation of the $\pi-$ and $\sigma-$ components of the line. Despite the relatively low magnetic field strength, this splitting is visible due to the high Land\'e factor of the line (see Table \ref{table:atomic_data}). The Land\'e factor of the other three lines is significantly lower, and for the light-bridge magnetic field strength they only show little splitting. All Stokes parameters were normalized for each map to the average continuum intensity in a quiet-Sun region. For the first and second maps (black and red dashed lines) the continuum is below unity, indicating that the temperature of the light bridge is lower than that of the quiet Sun. In the third map, the continuum is slightly above one. This is in agreement with the higher temperature found at that location in the GFPI data after the arrival of the plasma blob.

The linear polarization signals measured in Stokes $Q$ and $U$ perfectly coincide in the first and second maps. On the contrary, in the third map some differences are apparent. After the plasma blob reaches the light bridge, the Stokes $Q$ signal is higher in the four spectral lines. Stokes $U$ does not show strong variations, except for the \FeI\ 15662.0 \AA\ which also presents a larger profile. 

In the first and second maps, the Stokes $V$ signals of \FeI\ 1564.85, 1565.29, and 15662.0 \AA\ are very similar, whereas the third map exhibits a reduction in its amplitude. Interestingly, for the third map the Stokes $V$ signal of the \FeI\ 15648.5, 15652.9 and 15662.0 \AA\ lines shows a three-lobed profile. These lines have a significant Land\'e factor and are sensitive to the magnetic field. In addition to the regular antisymmetric Stokes $V$ profile,
the three-lobed signal shows another lower amplitude lobe at the red wing of the profile. This additional lobe is of the opposite polarity as the adjacent lobe. The Stokes $V$ profile of the \FeI\ 15648.5 \AA\ line apparently shows a fourth lobe at the blue wing. That bump in the profile is produced by the blend of the near \FeI\ 15647.4 \AA\ line \citep[\eg,][]{Ruedi+etal1995b}. Additional lobes are not seen in the lower magnetic sensitive \FeI\ 15665.2 \AA\ line. Three-lobed Stokes $V$ profiles are only found at an elongated region (with two pixels width) around the location of enhanced temperatures as inferred from the \FeI\ 6173 \AA\ inversions, that is, near the location where the plasma blob reaches the photosphere.

\subsection{Model selection and inversion}

The three-lobed and asymmetric Stokes parameters measured in the third GRIS map, after the arrival of the plasma blob, cannot be fitted with a simple model atmosphere. These complex profiles have been ascribed to the presence of gradients along the LOS in the magnetic and thermodynamic parameters of the atmosphere \citep[\eg,][]{Solanki+Pahlke1988, Sigwarth+etal1999, Khomenko+etal2003, Franz+Schlichenmaier2013} or to the presence of several atmospheric components in a single spatial resolution element \citep[\eg,][]{MartinezPillet+etal1994, Bernasconi+Solanki1996, Fischer+etal2012}. 

Both hypothesis apparently provide similarly good fits to the observed profiles. The preference of one model over the other is generally based on objective criteria. For example, amplitude asymmetries in Stokes $V$ can be reproduced by the presence of several unresolved magnetic atmospheres, with no need of including gradients in velocity or magnetic field. In contrast, correlated gradients between Doppler velocity and magnetic field along the LOS are required to produce net circular polarization (area asymetry). However, the selection of the complexity of the model or the number of free parameters is usually based on subjective assumptions, but this choice has a critical role in the results and interpretation of the data. In order to determine the model that best explains our observations, we calculated the Bayesian Information Criterion (BIC) for different fits following \citet{AsensioRamos+etal2012}. BIC is calculated following the expression

\begin{equation}
BIC=\chi_{\rm min}^2+k\ln N,
\label{eq:BIC}
\end{equation}

\noindent where $N$ is the number of data points (number of wavelength points multiplied by four for the four Stokes parameters), $k$ is the number of free parameters (the sum of the number of nodes for each variable), and $\chi_{\rm min}^2$ is obtained as

\begin{equation}
\chi_{\rm min}^2=\sum_{j=1}^N\Big (\frac{y_j(\hat{{\bf \theta}})-d_j}{\sigma_j}\Big )^2,
\label{eq:chi2}
\end{equation}

\noindent where $d_j$ are the measurements, $y_j$ are the model values for the set of parameters $\hat{{\bf \theta}}$ that minimize the $\chi^2$, and $\sigma_j$ is an estimation of the noise (determined from the standard deviation at the continuum). BIC provides a quantitative criterion for model comparison. It takes into account the quality of the fit, but penalizes an excess in the number of free parameters. 

For this analysis we focused on the individual inversion of the \FeI\ 15652.9 \AA\ line (instead of the simultaneous inversion of the four \FeI\ lines as done in Sect. \ref{sect:GRIS}) since it is isolated and does not show blends with other lines. The weight of Stokes $V$ was increased for the inversion. We evaluated the performance of three different models in fitting the Stokes parameters of that line: a single component atmosphere, a two-component model including a non-magnetic atmosphere, and a two magnetic components inversion. In all cases we systematically explored the values of BIC obtained for different number of free parameters by changing the number of nodes in the inversion. Table \ref{table:bic} shows the values of BIC for the cases with a variable number of nodes in velocity ($N_{\rm v}$), magnetic field strength ($N_{\rm B}$), and magnetic field inclination ($N_{\rm \gamma}$). All tabulated inversions used 5 nodes in temperature, 2 nodes in microturbulence, and 2 nodes in azimuth. Inversions with a different number of nodes in these quantities were also tested, but the results are not included in the table for simplicity. For the inversions with just one component we used as initial guess atmosphere the cool umbral model from \citet{Collados+etal1994} (similar to Sect. \ref{sect:GRIS}). In the case of the inversions with two components, the initial guess of the first component is the same umbral model (nodes are indicated by the columns) and the inversion of the same position of the light bridge following Sect. \ref{sect:GRIS} was selected as initial guess for the second component (nodes are indicated by the rows).

\begin{figure*}[!ht] 
 \centering
 \includegraphics[width=16cm]{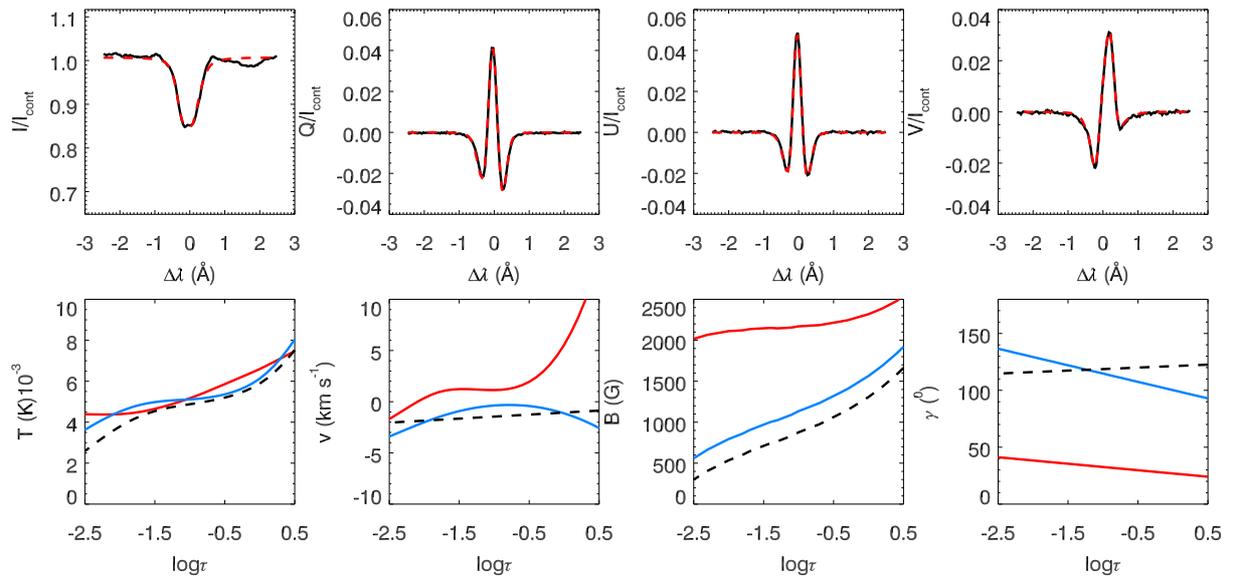}
  \caption{Top: Stokes profiles of the \FeI\ 15652.9 \AA\ line at the position indicated by the blue arrow in panel (a) of Fig. \ref{fig:time_GFPI} for the third GRIS map (black line) and the best fit obtained from a two magnetic components inversion with SIR (red dashed lines). Bottom: Stratification in temperature, Doppler velocity, magnetic field strength, and magnetic field inclination obtained at the same position from the first GRIS map (black dashed line), and the inversion of the third GRIS map for the first component (red, filling factor 0.15) and second component (blue, filling factor 0.85).}
  \label{fig:inversion}
\end{figure*}

The model with the lowest BIC is preferred. Its value is marked in bold in Table \ref{table:bic} and the corresponding fit and atmospheric stratification is illustrated in Fig. \ref{fig:inversion}. The analysis reveals that the most appropriate model for explaining the observation is the case with two magnetic components. The first component (red line in bottom panels of Fig. \ref{fig:inversion} and filling factor of 0.15) uses five nodes in velocity, and two nodes in magnetic field strength and inclination. The second component (blue line in bottom panels of Fig. \ref{fig:inversion} and filling factor of 0.85) uses the same number of nodes. Note that a difference in the BIC value between two models above 10 indicates a very strong evidence in favour of the model with lower BIC. Our study shows huge differences between any pair of models. All inversions with two magnetic components have a much lower BIC than any inversion with a single component or two-components with one of them non-magnetic. The analysis clearly points to a two magnetic component model as the most probable scenario. The atmosphere presents gradients in velocity and magnetic field, since otherwise it would not be able to reproduce the 8\% area asymetry of the Stokes $V$ profile.

\begin{table*}
\caption[]{\label{table:bic}
          {Bayesian Information Criterion for the inversion of the \FeI\ 15652.9 \AA\ line at the position indicated by a blue arrow in panel a of Fig. \ref{fig:time_GFPI} for the third GRIS map under different models.}}
     
\setlength{\aboverulesep}{0pt}
\setlength{\belowrulesep}{0pt}

\resizebox{6.5cm}{!} {      
          
\begin{tabular*}{10cm}{cccc?cc|cc|cc?cc|cc|cc?cc|cc|cc}


& & & &\multicolumn{6}{c?}{$N_{\rm v}$=2} & 
\multicolumn{6}{c?}{$N_{\rm v}$=5} & 
\multicolumn{6}{c}{$N_{\rm v}$=9}\\

& & & &\multicolumn{2}{c|}{$N_{\rm B}$=2} & 
\multicolumn{2}{c|}{$N_{\rm B}$=5} & 
\multicolumn{2}{c?}{$N_{\rm B}$=9} &

\multicolumn{2}{c|}{$N_{\rm B}$=2} & 
\multicolumn{2}{c|}{$N_{\rm B}$=5} & 
\multicolumn{2}{c?}{$N_{\rm B}$=9} &

\multicolumn{2}{c|}{$N_{\rm B}$=2} & 
\multicolumn{2}{c|}{$N_{\rm B}$=5} & 
\multicolumn{2}{c}{$N_{\rm B}$=9} \\

& & & & $N_{\rm \gamma}$=2 & $N_{\rm \gamma}$=5  & $N_{\rm \gamma}$=2 & $N_{\rm \gamma}$=5 & $N_{\rm \gamma}$=2 & $N_{\rm \gamma}$=5 & $N_{\rm \gamma}$=2 & $N_{\rm \gamma}$=5 & $N_{\rm \gamma}$=2 & $N_{\rm \gamma}$=5 & $N_{\rm \gamma}$=2 & $N_{\rm \gamma}$=5 & $N_{\rm \gamma}$=2 & $N_{\rm \gamma}$=5  & $N_{\rm \gamma}$=2 & $N_{\rm \gamma}$=5 & $N_{\rm \gamma}$=2 & $N_{\rm \gamma}$=5 \\ \cmidrule[2pt]{1-22}

1 comp & & & & {\tiny 15215} & {\tiny 5819} & {\tiny 17386} & {\tiny 7334} & {\tiny 17411}  & {\tiny 7359} & {\tiny 17067} & {\tiny 4869} & {\tiny 16953} & {\tiny 16343} & {\tiny 16978} & {\tiny 16368} & {\tiny 17092} & {\tiny 4894}  & {\tiny 16978} & {\tiny 16368} & {\tiny 17002} & {\tiny 16393} \\ \cmidrule[2pt]{1-22}

\multirow{3}{*}{\shortstack{2 comp \\ (non-mag.)}} & $N_{\rm v2}$=2 & & & {\tiny 18062} & {\tiny 6208} & {\tiny 18413} & {\tiny 6102} & {\tiny 18438}  & {\tiny 6127} & {\tiny 17287} & {\tiny 5166} & {\tiny 16898} & {\tiny 5456} & {\tiny 16923} & {\tiny 5481} & {\tiny 17312} & {\tiny 5191}  & {\tiny 16923} & {\tiny 5481} & {\tiny 16948} & {\tiny 5506} \\

			& $N_{\rm v2}$=5 & & & {\tiny 18067} & {\tiny 4173} & {\tiny 18466} & {\tiny 4843} & {\tiny 18490}  & {\tiny 4868} & {\tiny 17105} & {\tiny 6321} & {\tiny 16622} & {\tiny 6524} & {\tiny 16647} & {\tiny 6549} & {\tiny 17130} & {\tiny 6346}  & {\tiny 16647} & {\tiny 6549} & {\tiny 16672} & {\tiny 6574} \\

			& $N_{\rm v2}$=9 & & & {\tiny 18092} & {\tiny 4198} & {\tiny 18490} & {\tiny 4868} & {\tiny 18515}  & {\tiny 4892} & {\tiny 17130} & {\tiny 6346} & {\tiny 16647} & {\tiny 6549} & {\tiny 16672} & {\tiny 6574} & {\tiny 17155} & {\tiny 6371}  & {\tiny 16672} & {\tiny 6574} & {\tiny 16697} & {\tiny 6599} \\ \cmidrule[2pt]{1-22}

\multirow{18}{*}{\shortstack{2 comp \\ (mag.)}} & \multirow{6}{*}{$N_{\rm v2}$=2}  & \multirow{2}{*}{$N_{\rm B2}$=2} & $N_{\rm \gamma 2}$=2 & {\tiny 3138} & {\tiny 3161} & {\tiny 3059} & {\tiny 3084} & {\tiny 3084}  & {\tiny 3109} & {\tiny 3037} & {\tiny 3040} & {\tiny 2947} & {\tiny 2985} & {\tiny 2972} & {\tiny 3010} & {\tiny 3062} & {\tiny 3065}  & {\tiny 2972} & {\tiny 3010} & {\tiny 2997} & {\tiny 3035} \\

			 & & & $N_{\rm \gamma 2}$=5 & {\tiny 3111} & {\tiny 3089} & {\tiny 2974} & {\tiny 3060} & {\tiny 2998}  & {\tiny 3085} & {\tiny 3010} & {\tiny 3012} & {\tiny 2908} & {\tiny 2956} & {\tiny 2933} & {\tiny 2981} & {\tiny 3035} & {\tiny 3037}  & {\tiny 2933} & {\tiny 2981} & {\tiny 2958} & {\tiny 3006} \\
 \cmidrule[1pt]{3-22}
			 & & \multirow{2}{*}{$N_{\rm B2}$=5} & $N_{\rm \gamma 2}$=2 & {\tiny 3177} & {\tiny 3158} & {\tiny 3075} & {\tiny 3111} & {\tiny 3100}  & {\tiny 3136} & {\tiny 3078} & {\tiny 3119} & {\tiny 2994} & {\tiny 3017} & {\tiny 3019} & {\tiny 3041} & {\tiny 3103} & {\tiny 3144}  & {\tiny 3019} & {\tiny 3041} & {\tiny 3044} & {\tiny 3066} \\

			 & & & $N_{\rm \gamma 2}$=5 & {\tiny 3113} & {\tiny 3086} & {\tiny 2978} & {\tiny 3015} & {\tiny 3002}  & {\tiny 3040} & {\tiny 3034} & {\tiny 3057} & {\tiny 2999} & {\tiny 3005} & {\tiny 3024} & {\tiny 3030} & {\tiny 3059} & {\tiny 3082}  & {\tiny 3024} & {\tiny 3030} & {\tiny 3049} & {\tiny 3054} \\	
 \cmidrule[1pt]{3-22}			 
			 & & \multirow{2}{*}{$N_{\rm B2}$=9} & $N_{\rm \gamma 2}$=2 & {\tiny 3201} & {\tiny 3183} & {\tiny 3100} & {\tiny 3136} & {\tiny 3125}  & {\tiny 3161} & {\tiny 3103} & {\tiny 3144} & {\tiny 3019} & {\tiny 3041} & {\tiny 3044} & {\tiny 3066} & {\tiny 3128} & {\tiny 3169}  & {\tiny 3044} & {\tiny 3066} & {\tiny 3069} & {\tiny 3091} \\
			 
			 & & & $N_{\rm \gamma 2}$=5 & {\tiny 3138} & {\tiny 3111} & {\tiny 3002} & {\tiny 3040} & {\tiny 3027}  & {\tiny 3065} & {\tiny 3059} & {\tiny 3082} & {\tiny 3024} & {\tiny 3030} & {\tiny 3049} & {\tiny 3054} & {\tiny 3084} & {\tiny 3107}  & {\tiny 3049} & {\tiny 3054} & {\tiny 3074} & {\tiny 3079} \\ \cmidrule[1.5pt]{2-22}

			 & \multirow{6}{*}{$N_{\rm v2}$=5}  & \multirow{2}{*}{$N_{\rm B2}$=2} & $N_{\rm \gamma 2}$=2 &  {\tiny 3167}  & {\tiny 3099} & {\tiny 2970} & {\tiny 3018} & {\tiny 2995}  & {\tiny 3043} & {\tiny {\bf 2884}} & {\tiny 3073} & {\tiny 2939} & {\tiny 2899} & {\tiny 2964} & {\tiny 2909} & {\tiny 2924} & {\tiny 3097}  & {\tiny 2964} & {\tiny 2909} & {\tiny 2989} & {\tiny 2934} \\

			 & & & $N_{\rm \gamma 2}$=5 & {\tiny 2994} & {\tiny 3060} & {\tiny 2999} & {\tiny 3086} & {\tiny 3024}  & {\tiny 3111} & {\tiny 2992} & {\tiny 2982} & {\tiny 2950} & {\tiny 2981} & {\tiny 2974} & {\tiny 3006} & {\tiny 3017} & {\tiny 3007}  & {\tiny 2974} & {\tiny 3006} & {\tiny 2999} & {\tiny 3031} \\
 \cmidrule[1pt]{3-22}
			 & & \multirow{2}{*}{$N_{\rm B2}$=5} & $N_{\rm \gamma 2}$=2 & {\tiny 3165} & {\tiny 3095} & {\tiny 3076} & {\tiny 2989} & {\tiny 3101}  & {\tiny 3014} & {\tiny 2979} & {\tiny 3004} & {\tiny 2958} & {\tiny 2964} & {\tiny 2982} & {\tiny 2989} & {\tiny 3004} & {\tiny 3029}  & {\tiny 2982} & {\tiny 2989} & {\tiny 3007} & {\tiny 3014} \\

			 & & & $N_{\rm \gamma 2}$=5 & {\tiny 2962} & {\tiny 3080} & {\tiny 2949} & {\tiny 2970} & {\tiny 2974}  & {\tiny 2995} & {\tiny 2955} & {\tiny 2924} & {\tiny 2922} & {\tiny 2966} & {\tiny 2947} & {\tiny 2991} & {\tiny 2980} & {\tiny 2949}  & {\tiny 2947} & {\tiny 2991} & {\tiny 2972} & {\tiny 3016} \\	
 \cmidrule[1pt]{3-22}			 
			 & & \multirow{2}{*}{$N_{\rm B2}$=9} & $N_{\rm \gamma 2}$=2 & {\tiny 3190} & {\tiny 3120} & {\tiny 3101} & {\tiny 3014} & {\tiny 3126}  & {\tiny 3039} & {\tiny 3004} & {\tiny 3029} & {\tiny 2982} & {\tiny 2989} & {\tiny 3007} & {\tiny 3014} & {\tiny 3029} & {\tiny 3054}  & {\tiny 3007} & {\tiny 3014} & {\tiny 3032} & {\tiny 3039} \\
			 
			 & & & $N_{\rm \gamma 2}$=5 & {\tiny 2987} & {\tiny 3105} & {\tiny 2974} & {\tiny 2995} & {\tiny 2998}  & {\tiny 3020} & {\tiny 2980} & {\tiny 2949} & {\tiny 2947} & {\tiny 2991} & {\tiny 2972} & {\tiny 3016} & {\tiny 3004} & {\tiny 2973}  & {\tiny 2972} & {\tiny 3016} & {\tiny 2997} & {\tiny 3041} \\			 
			 \cmidrule[1.5pt]{2-22}

			 & \multirow{6}{*}{$N_{\rm v2}$=9}  & \multirow{2}{*}{$N_{\rm B2}$=2} & $N_{\rm \gamma 2}$=2 & {\tiny 3192} & {\tiny 3124} & {\tiny 2995} & {\tiny 3043} & {\tiny 3020}  & {\tiny 3068} & {\tiny 2924} & {\tiny 3097} & {\tiny 2964} & {\tiny 2909} & {\tiny 2989} & {\tiny 2934} & {\tiny 2949} & {\tiny 3122}  & {\tiny 2989} & {\tiny 2934} & {\tiny 3014} & {\tiny 2958} \\

			 & & & $N_{\rm \gamma 2}$=5 & {\tiny 3019} & {\tiny 3084} & {\tiny 3024} & {\tiny 3111} & {\tiny 3049}  & {\tiny 3136} & {\tiny 3017} & {\tiny 3007} & {\tiny 2974} & {\tiny 3006} & {\tiny 2999} & {\tiny 3031} & {\tiny 3042} & {\tiny 3032}  & {\tiny 2999} & {\tiny 3031} & {\tiny 3024} & {\tiny 3055} \\
 \cmidrule[1pt]{3-22}
			 & & \multirow{2}{*}{$N_{\rm B2}$=5} & $N_{\rm \gamma 2}$=2 & {\tiny 3190} & {\tiny 3120} & {\tiny 3101} & {\tiny 3014} & {\tiny 3126}  & {\tiny 3039} & {\tiny 3004} & {\tiny 3029} & {\tiny 2982} & {\tiny 2989} & {\tiny 3007} & {\tiny 3014} & {\tiny 3029} & {\tiny 3054}  & {\tiny 3007} & {\tiny 3014} & {\tiny 3032} & {\tiny 3039} \\

			 & & & $N_{\rm \gamma 2}$=5 & {\tiny 2987} & {\tiny 3105} & {\tiny 2974} & {\tiny 2995} & {\tiny 2998}  & {\tiny 3020} & {\tiny 2980} & {\tiny 2949} & {\tiny 2947} & {\tiny 2991} & {\tiny 2972} & {\tiny 3016} & {\tiny 3004} & {\tiny 2973}  & {\tiny 2972} & {\tiny 3016} & {\tiny 2997} & {\tiny 3041} \\	
 \cmidrule[1pt]{3-22}			 
			 & & \multirow{2}{*}{$N_{\rm B2}$=9} & $N_{\rm \gamma 2}$=2 & {\tiny 3215} & {\tiny 3145} & {\tiny 3126} & {\tiny 3039} & {\tiny 3151}  & {\tiny 3064} & {\tiny 3029} & {\tiny 3054} & {\tiny 3007} & {\tiny 3014} & {\tiny 3032} & {\tiny 3039} & {\tiny 3054} & {\tiny 3079}  & {\tiny 3032} & {\tiny 3039} & {\tiny 3057} & {\tiny 3064} \\
			 
			 & & & $N_{\rm \gamma 2}$=5 & {\tiny 3012} & {\tiny 3129} & {\tiny 2998} & {\tiny 3020} & {\tiny 3023}  & {\tiny 3045} & {\tiny 3004} & {\tiny 2973} & {\tiny 2972} & {\tiny 3016} & {\tiny 2997} & {\tiny 3041} & {\tiny 3029} & {\tiny 2998}  & {\tiny 2997} & {\tiny 3041} & {\tiny 3022} & {\tiny 3066} \\

 \cmidrule[2.5pt]{1-22}

\end{tabular*}

}

\begin{tablenotes}
\small
 \item

 Columns indicate the number of nodes in velocity ($N_{\rm v}$), magnetic field strength ($N_{\rm B}$), and inclination ($N_{\rm \gamma}$) for the component using the cool umbral model from \citet{Collados+etal1994} as initial guess. Rows correspond to models with one component (first row), two components including a non-magnetic atmosphere (second to fourth rows), and two magnetic components (fifth to last rows). Each row indicates the number of nodes in velocity ($N_{\rm v2}$), magnetic field strength ($N_{\rm B2}$), and inclination $N_{\rm \gamma 2}$ of the second component. The BIC value of the best model ($N_{\rm v}=5$, $N_{\rm B}=2$, $N_{\rm \gamma}=2$, $N_{\rm v2}=5$, $N_{\rm B2}=2$, $N_{\rm \gamma 2}=2$) is highlighted in bold.

\end{tablenotes}

\end{table*}

\section{Discussion}
\label{sect:discussion}

\subsection{Initial brightenings and flare triggering}

Figure \ref{fig:GRIS_inv} shows that the $LB_{\rm e}$ presents upflows in a region with low horizontal magnetic field, while a downflow is found in a narrower part of the light bridge, closer to the center of the umbra. The upflow constantly transports new magnetic flux (horizontal fields) in a region surrounded by stronger vertical fields belonging to the umbra. As a result, reconnection events can be triggered between horizontal and vertical fields \citep{Toriumi+etal2015b, Toriumi+etal2015a} producing some of the brightenings seen in G-band and \CaIIH\ images as well as photospheric heating, like reported in our observations between 09:05 and 09:33 UT (Fig. \ref{fig:time_GFPI}).

The arch filament system of the active region shows an interesting configuration, where the light bridge is connected by magnetic field lines to the origin of a C-class flare in another spot. Only a few works \citep[\eg,][]{Yang+etal2016} have previously reported this kind of structure, while a light wall (above a sunspot light-bridge) has been found to form a coupled system with a flare loop \citep{Hou+etal2016}. Our analysis of AIA filtergrams and LOS magnetograms is not enough to understand the relation between the flare and the light bridge. Computing magnetic field extrapolations from vector magnetograms from HMI of the whole active region at different stages of the flare would shed light on the behavior of the arch filament system, but this is beyond the scope of this paper. What we deduce from the HMI continuum images and magnetograms (see Fig. \ref{fig:evolution_HMI} and associated movie) is that the formation of the light bridge seems to coincide with the start of the shearing of the two polarities at the flaring location to the east. The umbra is slowly invading a protruding quiet penumbral region. We speculate that this is related to the shearing in the flaring site, where one polarity loops to the sunspot here. In addition, the shearing at the flare site seems to start about 10 hours before the flare, and it is exactly during this time the light bridge develops. The light bridge slowly disappears after the flare, \ie, after the shearing at the flare site ceased.

Interestingly, some brightenings and heating events are found at the exact location of one of the footpoints of the active region loop that links the light bridge with the area where the C-class flare started. The first of these brightenings occurs during the flare eruption. Just prior to the eruption ($t_{\rm f}=$09:47 UT) a small increase in the intensity is found in the G-band and \CaIIH\ images, in addition to a small temperature increase inferred from the \FeI\ 6173 \AA\ line. At the same time, a negative polarity region moves towards a positive polarity region (blue arrow in Fig. \ref{fig:evolution_HMI} and associated movie). Photospheric shearing motions produce changes in the above laying magnetic field lines, where one of the footpoints of the loop is located. The other footpoint is above the complex magnetic field topology of the light bridge. A small perturbation of the loop can result in magnetic reconnection in the light-bridge region, which is seen as brightenings and temperature increments close to the time of the flare eruption. Meanwhile, the shear below the other footpoint continues and the energy builds-up until the flare is finally triggered \citep[\eg][]{Shibata+Magara2011}. During the following 10 hours the two regions with opposite polarities perform a vortex-like motion, and the energy is successively stored and released as five C-class flares.

\subsection{Plasma velocity and heating}

We estimated the velocity of the plasma blob along the loop. Unfortunately, we were not able to determine the 3D configuration of the loop. We were unable to infer it from stereoscopic observations due to the unavailability of Solar Terrestrial Relations Observatory (STEREO) data, since both spacecrafts were located at the farside hemisphere of the Sun during the analyzed event. We were also unable to retrieve a reliable estimation from geometrical considerations based on the projected positions of the loop \citep[\eg,][]{Verwichte+etal2010}. The LOS is contained in the loop plane and, thus, the loop projection in the plane of the sky is independent on the height of the loop $(h)$. Instead, we have performed a rough estimation of the loop length and interpreted it in a conservative way.  

From AIA 304 \AA\ images, we measured a distance of 38 Mm  between the footpoints of the loop ($2a$, one of them located at the flare origin and the other one above the light bridge). The plasma blob expends 360 s traveling between the mid point of the loop and the footpoint above the light bridge. The arrival time to the footpoint is defined as the time when the emission of the plasma blob in AIA 304 \AA\ completely vanishes. Neglecting the curvature of the loop, the average velocity of the plasma is \hbox{$\sim53$ km s$^{-1}$}, while the assumption of an elliptical loop with a ratio $h/a=0.5$ yields \hbox{$\sim66$ km s$^{-1}$}. These values provide an estimation of the minimum velocity of the plasma blob moving along the active region loop. They are in agreement with the velocities found in numerical simulations by \citet{Zacharias+etal2011}.

The impact of the blob on the light bridge produces localized brightenings in G-band and \CaIIH, in addition to photospheric heating (Fig. \ref{fig:time_GFPI}). The intensity emission of the \CaIIH\ line arises from a wide range of atmospheric heights, between the continuum forming layer and the lower chromosphere \citep{Vernazza+etal1981, Carlsson+Stein1997}. In this region the sound speed is around 10-20 km s$^{-1}$, much lower than the minimum average velocity of the plasma blob. Since the Doppler velocity inferred from the \FeI\ 6173 \AA\ line does not show unusually high speeds, the plasma must undergo a strong deceleration before reaching the photosphere. This process may occur as the fast moving plasma enters the low chromosphere (where the sound speed is lower) and develops into a supersonic shock whose velocity then is reduced to subsonic speed. The energy dissipated by the shock can produce the observed increase in emission \citep[\eg,][]{Cargill+Priest1980}. This energy input can also trigger new reconnection events at the light bridge, which also contribute to the emission and heating at lower atmospheric layers.

\citet{Toriumi+etal2017b} observed heating events produced by falling material along arch filaments in emerging flux regions. Their reported supersonic downflow velocities are consistent with those presented in this work. From IRIS observations, \citet{Kleint+etal2014} detected much stronger supersonic downflows (up to 200 km s$^{-1}$) associated to brightenings in the transition region above sunspots. They appear at the footpoints of the active region loops, and are the result of cool plasma falling from coronal heights, such as coronal rain. Using also IRIS data, \citet{Tian+etal2014} reported bright dots located mainly in the penumbra moving in the radial direction with speeds below 40 km s$^{-1}$. These features were interpreted as reconnection in the transition region and chromosphere, although the authors pointed out that some of the dots could be associated with falling plasma.

\subsection{Impact on the light-bridge atmosphere}

In the third GRIS map, the Stokes parameters at the impact region of the blob show significant differences with respect to those prior to the arrival of the plasma. Their main characteristic is the presence of three-lobed Stokes $V$ profiles in the three magnetically sensitive infrared \FeI\ lines. We proved that these signals have to be interpreted as two-component atmosphere inside the resolution element, whose second component appears as a result of the new plasma that reaches the photospheric light bridge.     

Most of the resolution element is filled with an atmosphere (blue lines in bottom panels of Fig. \ref{fig:inversion}) similar to that inferred for the unperturbed light bridge (black dashed lines in bottom panels of Fig. \ref{fig:inversion}). At the photosphere, it has a mostly horizontal magnetic field with a strength of around 1500 G, a few hundred Gauss higher than that obtained from the first map. Its Doppler velocity indicates upflows with a similar magnitude of that measured before the impact of the plasma blob, while the temperature stratification shows a slightly higher photospheric temperature. This region (between $\log\tau =0$ and $\log\tau =0.5$) corresponds to the highest sensitivity of the \FeI\ 15652.9 \AA\ line to the temperature \citep[\eg,][]{Borrero+etal2016}. As previously discussed, the atmosphere was heated after the impact of the hot plasma blob. 

The component with lower filling factor (red lines in bottom panels of Fig. \ref{fig:inversion}) shows  downflows of around 5 km s$^{-1}$ at the low photosphere. We conjecture that this component belongs to the falling blob. Its velocity is below the average velocity estimated for the travel of the plasma blob, indicating that it was decelerated at higher atmospheric layers. The magnetic field of this component has a stronger field strength and a lower inclination than the original magnetic field at the light bridge. Interestingly, its polarity is opposite to that of the umbra. We speculate on two different scenarios which can conduct to this finding. On the one hand, this magnetic topology may be the direct result of magnetic reconnection taking place due to the action of the plasma blob on the complex magnetic field topology of the light bridge. At $\log\tau=-0.3$, this component of the inversion shows an increase of the temperature compared to the main component, which may be associated to energy released during the reconnection event. On the other hand, the supersonic falling of the plasma blob can push down the light-bridge material and reverse the overlying canopy field lines, producing a polarity opposite to the umbra. \citet{Felipe+etal2016b} reported field reversals produced by the action of the convective motions in the low magnetized light-bridge atmosphere. In this scenario, the temperature increase is caused by the shock, as the plasma undergoes a velocity reduction from supersonic to subsonic speeds.

\section{Conclusions}
\label{sect:conclusions}

Based on multi-instrument observations, we have reported photospheric and chromospheric brightenings, heating events, and changes in the magnetic field produced by the impact of plasma accelerated by a C-class flare on a sunspot light-bridge. The plasma blob was originated during the moderately energetic flare and remained confined within an active region loop. One of the footpoints of the loop was located above the light bridge. 

We analyzed the photospheric magnetic and thermal structure of a sunspot light bridge before, during, and after the occurrence of a nearby flare. Our results match common properties of light bridges, such as weaker and more horizontal magnetic fields with respect to the surrounding umbral atmosphere. More interestingly, we measured several signatures related to the impact of the plasma blob on the light bridge. It is well known that field lines above the light bridge form a canopy structure, where vertical field lines coming from both sides of the light bridge merge at the top \citep{Jurcak+etal2006, Lagg+etal2014, Felipe+etal2016b}. Our observations show that one of the footpoints of the active region loop is located just above this canopy, and connect the upper atmosphere of the light bridge with the origin of the flare. 

The perturbation by the flare of the magnetic field of the loop can trigger small reconnection events at the light bridge, due to the interaction of its weak horizontal field with the surrounding vertical field. The reconnection manifests as brightenings in the G-band and \CaIIH\ images and as a photospheric temperature increment. The subsequent impact of the plasma ejected from the flare on the light bridge is associated with enhanced brightenings and heating at the photosphere and low chromosphere. Its trace was measured in the Stokes profiles, which were interpreted as arising from a two-component atmosphere inside the resolution element. The dominant component accounts for the heated but otherwise unperturbed light-bridge atmosphere, while the other component includes the contribution of the falling plasma blob and possibly the reconnection event triggered by its impact.

\begin{acknowledgements} 
The 1.5-meter GREGOR solar telescope was built by a German consortium under the
leadership of the Kiepenheuer-Institut f\"ur Sonnenphysik in Freiburg with the
Leibniz-Institut f\"ur Astrophysik Potsdam, the Institut f\"ur Astrophysik
G\"ottingen, and the Max-Planck-Institut f\"ur Sonnensystemforschung in G\"ottingen as
partners, and with contributions by the Instituto de Astrof\'isica de Canarias and
the Astronomical Institute of the Academy of Sciences of the Czech Republic. SDO data are provided by the Joint Science Operations Center - Science Data Processing. The support of the European Commission's FP7 Capacities Programme under Grant Agreement number 312495 ``SOLARNET'' is acknowledged. This research has been funded by the Spanish MINECO through grants AYA2014-55078-P, AYA2014-60476-P and AYA2014-60833-P.
\end{acknowledgements}

\bibliographystyle{aa} 
\bibliography{biblio.bib}

\end{document}